\documentclass[journal=jctcce,manuscript=article,layout=twocolumn]{achemso}
\setkeys{acs}{articletitle = true}

\usepackage{graphicx}
\usepackage{amsmath}
\usepackage{txfonts}
\usepackage{hyperref}
\usepackage{xcolor}
\usepackage{multirow}
\usepackage{booktabs}
\usepackage{CJKutf8}
\usepackage{xr}
\usepackage[version=4]{mhchem}

\title{Converging TDDFT calculations in 5 iterations with minimal auxiliary preconditioning}
\author{Zehao Zhou} 
\affiliation{Department of Frontier Key Technology, Changping Laboratory \\ 28 Kexueyuan Road, Beijing, 102206, China}
\author{Shane M. Parker}
\email{shane.parker@case.edu}
\affiliation{Department of Chemistry, Case Western Reserve University \\ 10900 Euclid Ave, Cleveland, OH 44106, USA}

\begin{document}

\begin{abstract}
Eigenvalue problems and linear systems of equations involving large symmetric matrices
are commonly solved in quantum chemistry using Krylov space methods, such as the Davidson
algorithm. The preconditioner is a key component of Krylov space methods that accelerates
convergence by improving the quality of new guesses at each iteration.
We systematically design a new preconditioner for time-dependent density functional theory
(TDDFT) calculations based on the recently introduced TDDFT-ris semiempirical model by
re-tuning the empirical scaling factor and the angular momenta of a minimal
auxiliary basis.
The final preconditioner produced includes up to $d$-functions in the auxiliary basis
and is named ``rid''.
The rid preconditioner converges excitation energies and polarizabilities in 5-6
iterations on average, a factor of 2-3 faster than the conventional diagonal
preconditioner, without changing the converged results.
Thus, the rid preconditioner is a broadly applicable and efficient preconditioner
for TDDFT calculations.
\end{abstract}

\section{Introduction}
Eigenvalue problems and linear systems of equations involving large symmetric matrices are
ubiquitous in quantum chemistry. By large, we refer to matrices
that are too large to be explicitly stored in memory or on disk.
Such problems appear as rate-limiting steps in a variety of
contexts.
For example,
correlation energies within configuration interaction methods are obtained
as eigenvalues of a large Hamiltonian matrix with dimensions of up to
trillions of Slater determinants.\cite{Siegbahn1977CP,Knowles1984CPL,Olsen1990CPL,Mitrushenkov1994CPL,Gao2024JCTC}
Within response theory, excitation energies are obtained
as eigenvalues of a response operator and (non)linear properties require
the solution of linear systems of equations with respect to the response matrix;
this response matrix has linear dimensions that scale at the least as
$\mathcal{O}(N^2)$ where $N$ is some measure of system size, such that storage
of the full response matrix would require at least $\mathcal{O}(N^4)$ storage.
\cite{Olsen1985JCP,Christiansen1998IJQC,Parker2018FoQC}
Similarly, stability analyses of self-consistent-field
solutions reduce to computing eigenvalues of orbital rotation
Hessians.\cite{Thouless1960NP,Cizek1967JCP,Bauernschmitt1996JCP}

The Krylov subspace approach is one of the most widely used
and most successful strategies for iteratively computing a few extremal
eigenpairs or solving linear equations in quantum chemistry.\cite{davidson1975JCoP}
Methods based on Krylov subspace
approaches forego the need to store large matrices by centering the algorithm
on matrix-vector products, $\mathbf{A}\mathbf{V}$, which can be computed on-the-fly
and can be highly optimized for specific methods.\cite{davidson1975JCoP}
The eigenpairs are then written in terms of a small subspace, referred to as the
Krylov subspace, which is expanded in each iteration.
For most applications in quantum chemistry, the largest bottleneck by far is
the calculation of matrix-vector products. Hence, significant effort has gone
into accelerating these matrix-vector product routines, including through
screening techniques,\cite{Weiss1993JCP,furche2016JCP,Parrish2016JCTC}
tensor decomposition techniques,\cite{Bauernschmitt1997CPL,Hu2020JCTC}
and hardware acceleration.\cite{Ufimtsev2008JCTC,Isborn2011JCTC}

The number of matrix-vector products needed is determined by how
the Krylov subspace is expanded in each iteration. This step is referred to as
the preconditioning step, and the efficiency of a Krylov subspace method is
largely determined by the efficiency of the preconditioner.
However, comparatively less effort has gone into designing efficient
preconditioners.
The major advantage to improving the preconditioner is that it does not change
the final converged result of the Krylov subspace method. In addition, it
remains compatible with other approaches to accelerating the matrix-vector
product routines mentioned above. Thus, in this paper, we will focus on
accelerating the Krylov subspace method by designing powerful preconditioners.

For quantum chemistry applications, we seek preconditioners that
i) are significantly less
expensive to apply than the original matrix to ensure that the overall
computational cost is reduced, and ii) do not require system-specific tuning
(although they will inevitably be method specific).
In our previous research, we showed that semiempirical models are
attractive preconditioners because they are typically several orders of
magnitude cheaper than corresponding ab initio methods, can be broadly
applicable, and often contain the most essential ingredients of the underlying
physics.\cite{zhou2021JCP} We demonstrated this by using the semiempirical
simplified Tamm-Dancoff\cite{grimme2013JCP} (sTDA) and simplified
time-dependent density functional theory (sTDDFT)
models as preconditioners for computing excitation energies
and polarizabilities with ab initio TDDFT, which led to a factor of 1.6
speed up on average for excitation energies and a factor of 1.2 speed up
on average for polarizabilities.\cite{zhou2021JCP}
Although the speedup of excitation energies with sTDA/sTDDFT is already
significant, we believe the modest speedup of the polarizabilities suggests
that further improvements are possible with a semiempirical model that more
accurately reproduces transition densities.

In this paper, we systematically design a preconditioner for TDDFT excitation
energies and polarizability calculations based on the
minimal auxiliary basis approach for TDDFT, TDDFT-ris, recently
introduced by us.\cite{zhou2023JPCL} The TDDFT-ris model has the same
basic structure as the sTDA model, but
significantly outperforms the sTDA model in the accuracy of the excitation
energies, with just 0.06 eV error relative ab initio TDDFT compared to
0.24 eV error for the sTDA model. In addition, the TDDFT-ris model
shows exceptional accuracy in the UV-vis absorption spectra for small to
medium-sized organic molecules, indicating the oscillator strengths (and hence
transition densities) are more accurately captured than in sTDA/sTDDFT.
Furthermore, the TDDFT-ris model has additional flexibility in its structure
that enables a greater degree of design than would be possible with the sTDA
model.

Thus, on account of its superior accuracy and flexibility,
we anticipate that TDDFT-ris will
be a powerful preconditioner for TDDFT excitation energies and polarizability
calculations. The final preconditioner proposed in this paper, termed rid,
converges excitation energies in 5-6 iterations on average and converges linear
equations in about 6 iterations on average, compared to 12-17 iterations for
excitation energies and 12-13 iterations for linear equations using the
conventional diagonal preconditioner.
Furthermore, the rid preconditioner all but erases the difference in the
number of iterations needed to converge excitation energies with global
hybrid vs range-separated hybrid functionals.
We note that although it has been suggested that improving the
preconditioner can paradoxically worsen the convergence of the
algorithm,\cite{Sleijpen1996SJMAA,Hochstenbach2006G} we observe no
such deterioration in practice.

This paper is organized as follows: In section
\ref{sec:theory}, we briefly review the
preconditioned Krylov subspace algorithm,
as well as the working equations for ab initio TDDFT and the
semiempirical TDDFT models used in this work.
In section \ref{sec:design}, we systematically design the rid preconditioner.
In section \ref{sec:results}, we evaluate the rid preconditioner by comparing its
performance to the diagonal preconditioner and sTDA preconditioner when
computing the excitation energies or polarizability. Finally,
we conclude in section \ref{sec:conclusion} by offering our perspective on
how to make broadly applicable preconditioners for Krylov subspace methods.


\section{Theory and Methods}\label{sec:theory}
\subsection{Krylov Subspace Methods}
Here, we briefly describe the framework of the semiempirical preconditioned
Krylov subspace method, taking the Davidson algorithm as a representative
example.\cite{davidson1975JCoP}
The goal of the Davidson algorithm is to find the $N_\text{states}$
lowest eigenvalues and corresponding eigenvectors
of a Hermitian matrix $\mathbf{A}$.
The Davidson algorithm works by iteratively expanding a subspace
(called the Krylov subspace) spanned by a set of basis vectors
until it contains the desired eigenpairs to a specified threshold.
At the $k$-th iteration, the Krylov subspace is spanned by the
columns of the matrix $\mathbf{V}^{(k)}$. Here, we assume
the columns of $\mathbf{V}^{(k)}$ form an orthonormal basis, but that
is not required.\cite{furche2016JCP}
To accelerate convergence, the Davidson algorithm makes use of
an approximation to $\mathbf{A}$---called the preconditioner and
denoted here using $\mathbf{T}$---which will be used to
i) generate the initial subspace, $\mathbf{V}^{(0)}$,
and ii) generate expansion directions at each iteration.
In conventional implementations, $\mathbf{T}=\mathbf{A}^\text{diag}$
is the diagonal approximation of $\mathbf{A}$.
In this paper we will consider semiempirical models
for $\mathbf{T}$.

The Davidson algorithm begins by generating the initial subspace,
usually by using the low-lying eigenvectors of $\mathbf{T}$.
In principle, so long as the true eigenvectors have a nonzero overlap
with the initial subspace, then the Davidson algorithm will arrive at the
correct solution. To help ensure this, additional initial vectors are
often included in the initial subspace, i.e., $N_\text{init} > N_\text{states}$,
where $N_\text{init}$ is the number of initial guesses.
Regardless of how many initial vectors are used,
in each iteration, the subspace projection of $\mathbf{A}$ is computed
as
\begin{equation}
  \mathbf{a}^{(k)} = \mathbf{V}^{(k),\dagger} \mathbf{A} \mathbf{V}^{(k)}.
\end{equation}
The approximate eigenvalues are obtained from $\mathbf{a}^{(k)}$ by diagonalizing,
\begin{equation}
  \mathbf{a}^{(k)}\mathbf{x}_n^{(k)} = \Omega_n^{(k)} \mathbf{x}_n^{(k)},
\end{equation}
and the approximate eigenvectors are obtained from
\begin{equation}
  \mathbf{X}^{(k)}=\mathbf{V}^{(k)}\mathbf{x}^{(k)}.
\end{equation}

\begin{figure}[tbp]
  \centering
  \includegraphics[width=\linewidth]{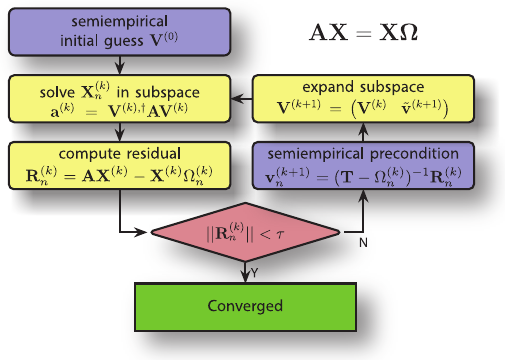}
  \caption{\label{fig:flow_chart}
  The scheme of Davidson algorithm with semiempirical preconditioning.
  }
\end{figure}

Next, for each root, the residual is computed according to
\begin{equation} \label{eq:residual}
  \mathbf{R}_n^{(k)}=\mathbf{A}\mathbf{X}^{(k)}_n - \mathbf{X}^{(k)}_n \Omega_n^{(k)}.
\end{equation}
If the residual norm, $|R_n^{(k)}|$, is smaller than a user-defined threshold
then that eigenpair is considered converged. For each unconverged eigenpair,
the residual is preconditioned to generate new search directions,
\begin{equation}
\label{eq:precondition}
\mathbf{v}_n^{(k+1)}=(\mathbf{T}-\Omega_n^{(k)})^{-1}\mathbf{R}_n^{(k)}.
\end{equation}
Thus, the $\mathbf{T}$ should be chosen such that applying a shifted inverse
is inexpensive relative to matrix multiplications of $\mathbf{A}$.
When $\mathbf{T}$ is a diagonal matrix, the preconditioning step
reduces to element-wise division.
When $\mathbf{T}$ is a semiempirical model, we solve Eq. \eqref{eq:precondition}
using an inner Krylov subspace method.
Finally, the new search directions are orthogonalized against $\mathbf{V}^{(k)}$
and each other using the modified Gram-Schmidt procedure,
and then appended to $\mathbf{V}^{(k)}$ to form $\mathbf{V}^{(k+1)}$.
The algorithm is depicted schematically in Fig. \ref{fig:flow_chart}.

It has been argued that the step in Eq. \eqref{eq:precondition} should not
be considered a preconditioning step because if we replace $\mathbf{T}$ with
$\mathbf{A}$, then no update vector is produced.\cite{Sleijpen1996SJMAA}
This can be seen by inserting Eq. \eqref{eq:residual} into \eqref{eq:precondition},
which gives $\mathbf{v}_n^{(k+1)}=\mathbf{X}_n^{(k)}$.
This result has been used to argue that improving the preconditioner in
the Davidson algorithm can lead to stagnation, not
acceleration.\cite{Sleijpen1996SJMAA,Hochstenbach2006G,Olsen1990CPL}
Indeed, the Jacobi-Davidson algorithm was proposed to avoid the stagnation
of the Davidson algorithm.\cite{Sleijpen1996SJMAA,Hochstenbach2006G}
However, despite these strong theoretical arguments, observation of this stagnation
in practical quantum chemistry calculations remains
elusive.\cite{VanDam1996JCC,Rappoport2023JCC}
The solution to this apparent paradox comes from
the choice to initiate the algorithm with an eigenvector of $\mathbf{T}$, which
has two consequences for the stability of the algorithm. First, somewhat trivially,
if $\mathbf{T}$ was indeed identical to $\mathbf{A}$, then the residual would
be zero in the first step and there would be no need to precondition. Second,
including low-lying eigenvectors of $\mathbf{T}$ in the initial subspace guarantees
that the $\mathbf{T} - \Omega_n^{(k)}$ is not singular, and leads to nearly
identical behavior between the Davidson and Jacobi-Davidson algorithms in
practice.\cite{Notay2004SJMAA}

\subsection{Time-dependent density functional theory}
\label{sec:TDDFT}
For the remainder of this paper, labels $i,j$ will be used to denote
occupied orbitals, and $a,b$ will be used to denote virtual orbitals,
and $p,q,r,s$ will be used to denote generic orbitals.

\subsubsection{ab initio TDDFT}

Excitation energies within linear response TDDFT are found by
solving the symplectic eigenvalue problem
\begin{equation}
  \label{eq:Casida}
    \begin{pmatrix}
    \mathbf{A} & \mathbf{B} \\
    \mathbf{B} & \mathbf{A}
    \end{pmatrix}
    \begin{pmatrix}
    \mathbf{X}_n\\
    \mathbf{Y}_n
    \end{pmatrix}
    =
    \Omega_n
    \begin{pmatrix}
    \mathbf{1} & \mathbf{0} \\
    \mathbf{0} & \mathbf{-1}
    \end{pmatrix}
    \begin{pmatrix}
    \mathbf{X}_n\\
    \mathbf{Y}_n
    \end{pmatrix}
    ,
  \end{equation}
where
\begin{subequations} \label{eq:hessian}
\begin{align}
    (A + B)_{ia,jb} =& (\varepsilon_{a} - \varepsilon_i) \delta_{ab}\delta_{ij} + 2f^\text{xc}_{iajb} + 2(ia|jb) \nonumber\\
       & - c_x[(ib|ja)+(ij|ab)],\\
    (A - B)_{ia,jb} =& (\varepsilon_{a} - \varepsilon_i) \delta_{ab}\delta_{ij}  + c_x[(ib|ja)-(ij|ab)],
\end{align}
\end{subequations}
are the electric and magnetic orbital rotation Hessians, respectively, $\varepsilon_p$ is
the Kohn-Sham eigenvalue associated with Kohn-Sham orbital $\phi_p$,
$f^\text{xc}_{iajb}$ is a matrix element of the exchange-correlation kernel,
\begin{equation}
\label{eq:pqrs}
  (pq|rs) = \iint dx dx' \phi_p(x)\phi_q(x)\frac{1}{|x-x'|}\phi_r(x')\phi_s(x')
\end{equation}
is an electron repulsion integral (ERI), and $c_x$ denotes the Hartree--Fock
mixing coefficient. The eigenvectors, $|X_n,Y_n\rangle$, represent
transition densities.

In this paper, we focus on excitation energies within the Tamm--Dancoff
Approximation, which is obtained by setting $\mathbf{B}=0$ in the above,
thus reducing to the Hermitian eigenvalue problem
\begin{equation}
\label{eq:TDA}
  \mathbf{A} \mathbf{X}_n^\text{TDA} = \Omega_n^\text{TDA} \mathbf{X}_n^\text{TDA}.
\end{equation}

The TDDFT dynamic polarizability at frequency $\omega$ is computed as
\begin{equation}
  \label{eq:TDDFT_polar}
    \alpha_{vu}^\text{TDDFT}(\omega) =
    \begin{pmatrix}
    \boldsymbol{\mu}^v_\text{vo}\\
    \boldsymbol{\mu}^v_\text{ov}
    \end{pmatrix}^\dagger
    \left[
    \begin{pmatrix}
      \mathbf{A} & \mathbf{B} \\
      \mathbf{B} & \mathbf{A}
    \end{pmatrix}
    -
    \omega
    \begin{pmatrix}
      \mathbf{1} & \mathbf{0} \\
      \mathbf{0} & \mathbf{-1}
    \end{pmatrix}
    \right]^{-1}
    \begin{pmatrix}
      \boldsymbol{\mu}^u_\text{vo}\\
      \boldsymbol{\mu}^u_\text{ov}
    \end{pmatrix}
    ,
\end{equation}
where the $\boldsymbol{\mu}_\text{vo}^v$ and $\boldsymbol{\mu}_\text{ov}^v$
are the virtual-occupied and occupied-virtual blocks
matrix representation of the dipole moment
operator in the $v$ direction, respectively.


\subsubsection{Semiempirical TDDFT models} \label{sec:semiempiricals}
In recent years several semiempirical models have been developed to approximate
TDDFT that retain the ab initio Kohn-Sham reference
but use semiempirical approximations to the linear response function.
Here, we focus on the simplified
TDA/TDDFT\cite{grimme2013JCP,bannwarth2014CTC} (sTDA/TDDFT) and the TDDFT-ris
models as these were explicitly designed for use with hybrid density functionals.
The results here are expected to apply to semiempirical models
that focus on semilocal density functionals as well, such as
TDDFT+TB\cite{asadi-aghbolaghi2020JPCC} and TDDFT-as.\cite{giannone2020JCP}

TDDFT-ris and sTDA are both constructed similarly by
i) neglecting the exchange-correlation kernel, i.e., setting $f^\text{xc}=0$,
and ii) approximating the 4-index ERIs
with a contraction of 2-index and 3-index tensors in Eq. \eqref{eq:pqrs},
\begin{equation}
  \label{eq:pqrs_approx}
    (pq|rs) \approx \sum_{PQ} K_{pq}^P J_{PQ} K_{rs}^Q.
\end{equation}
It is the definition of these three tensors that
distinguishes the TDDFT-ris and sTDA.

In TDDFT-ris, ERIs are approximated using the resolution of the identity
(RI) approximation with a minimal auxiliary basis such that
\begin{equation} \label{eq:TDDFT-ris_approx}
  K_{pq}^P = (pq|P), \quad J_{PQ} = (P|Q)^{-1},
\end{equation}
where $P$, $Q$ label auxiliary basis functions and $(pq|P)$ and $(P|Q)$ are the
three-center and two-center ERIs,
respectively.\cite{baerends1973CP,dunlap1979JCP,eichkorn1995CPL,Bauernschmitt1997CPL,heinze2000JCP,neese2002CPL,pedersenthomasbondo2009TCA,weigend2009JCP,stoychev2017JCTC}
The auxiliary basis functions are Gaussians with exponents defined by the
element type, $A$, as
\begin{equation}
  \alpha_A = \frac{\theta}{R_A^2},
\end{equation}
where $R_A$ is the atomic radius tabulated by Ghosh et al,\cite{ghosh2008JMS}
and $\theta$ is a global scaling factor.
Previously, we chose $\theta=0.2$ by minimizing the error
in excitation energies and absorption spectra.\cite{zhou2023JPCL}
In our previous work, we exclusively used $s$-type functions in the
auxiliary basis, but here we will explore the impact of including
higher angular momentum functions as well.

In sTDA, the ERIs are approximated using parametrized transition monopoles with
\begin{equation}
  K_{pq}^A = \sum_{\nu \in A} C'_{\nu p}C'_{\nu q},
  \quad J_{AB} = \Gamma_{AB}
\end{equation}
where $\nu\in A$ indicates atomic orbitals centered on atom $A$,
$\mathbf{C}'$ is the L\"{o}wdin orthogonalized molecular orbital coefficient matrix,
and $\Gamma_{AB}$ is a damped Coulomb operator between atoms $A$ and $B$
of the form
  \begin{equation}
  \Gamma_{AB} = \left((R_{AB})^{y} + (c_y \eta_{AB})^{-y}\right)^{-\frac{1}{y}}
\end{equation}
where $R_{AB}$ is the distance between the two atoms $A$ and $B$, $y$ is
an empirically tuned parameter, $c_y$ is either 1 or the HFX mixing coefficient,
and $\eta_{AB}$ is the average chemical hardness. Separate $\Gamma$ functions
are used for the Coulomb and exchange terms with independently tuned empirical
parameters. Here we only use the default suggested parameters for the
density functionals used.\cite{grimme2013JCP,Risthaus2014PCCP}
To be clear, by sTDA/sTDDFT model here, we specifically refer to the
approximation to the
$\mathbf{A}$ and $\mathbf{B}$ matrices used in sTDA/sTDDFT. We note that
the sTDA method as implementd in the sTDA program includes additional
innovations, such as a configuration state function selection algorithm
and perturbative corrections.\cite{grimme2013JCP}

\subsection{Implementation details} \label{sec:implementation}
We implemented the algorithms described above in a
pilot python code. In our implementation,
the ground-state Kohn-Sham reference,
matrix-vector product subroutines with density fitting, and
all ERIs are provided by PySCF 2.3.0.\cite{sun2018WCMS} The
matrix-vector products for sTDA and ris-type preconditioners
are implemented in Python.\cite{zhou2023JPCL}

\section{Systematic Design of the rid Preconditioner}
\label{sec:design}

One of the strong advantages of the TDDFT-ris method over competing
semiempirical methods is that it can be straightforwardly extended to include
atomic multipoles by adding higher angular momentum functions (e.g., $p$ and
$d$ functions) to the auxiliary
basis.\cite{zhou2023JPCL}
In this section, we exploit the flexibility of the
TDDFT-ris model to systematically design an optimal TDDFT preconditioner. The
TDDFT-ris model has two essential design choices: i) the global scale factor
$\theta$ used to determine the exponents for the auxiliary basis, and ii) the
maximum angular momentum used in the auxiliary basis.
In our previous work, we chose
$\theta=0.2$ and used only $s$-type functions.
However, we do not expect the same parameters to
necessarily be optimal for preconditioning as well.
Furthermore, we expect
these two design choices to be correlated.
For example, the appearance of an optimal
$\theta$ was previously attributed to engineered error cancellation between the
approximated Coulomb contribution and the neglected exchange-correlation
kernel.\cite{zhou2023JPCL} Thus, improving the description of the
Coulomb energy may require a different $\theta$ to maintain the error cancellation.
Finally, we have the additional constraint that the semiempirical $\mathbf{T}$ must
remain significantly less expensive to apply than $\mathbf{A}$
so that the computational time saved by preconditioning is not simply
spent in the preconditioner instead.

We proceed by first exploring strategies to contain the cost of the
preconditioners without sacrificing efficacy. Next, we will explore
the performance of preconditioners with different maximum angular
momenta, while tuning $\theta$ within the range of 0.1--10.
As a benchmark, we compute the lowest 5 excited states using the PBE0
density functional\cite{perdew1996JCP} with the def2-SVP basis
set\cite{weigend2005PCCP} for the TUNE8 set, which is a subset of
the PRECOND19 set defined previously\cite{zhou2021JCP}
and repeated in the Supporting Information.
We choose a
convergence threshold of $10^{-5}$ for the residual norm.
As a figure of merit in this section we focus on the
average number of ab initio matrix-vector products, $N_\text{mv}$.

\subsection{Controlling the preconditioning cost} \label{sec:cost}
To avoid the preconditioner becoming the bottleneck of the algorithm,
we target a preconditioning cost on the order of 1-2\% of the overall
wall time, meaning that one matrix-vector multiplication using $\mathbf{T}$
needs to be about $10^3$ times faster than a matrix-vector multiplication
using $\mathbf{A}$.
We focus our efforts on the exchange terms because
for both TDDFT-ris and sTDA, the exchange contribution of
the matrix-vector products scale
as $\mathcal{O}(N^4)$, where $N$ is the number of basis functions.
We introduce an energy cutoff for the
exchange ERIs, $t_\text{cut}$, and neglect all ERIs
involving virtual molecular orbitals (MOs) higher in energy than
$\epsilon_\text{HOMO}+t_\text{cut}$ or occupied MOs lower in energy
than $\epsilon_\text{LUMO} - t_\text{cut}$,
where HOMO stands for highest occupied molecular
orbital and LUMO, lowest unoccupied molecular orbital.
We find that $t_\text{cut} = 40$ eV allows us to dramatically reduce
the cost of applying $\mathbf{T}$ without
significantly increasing the $N_\text{mv}$.
We do not truncate any MOs for the Coulomb terms because they are
much less expensive than the exchange terms, scaling as $\mathcal{O}(N^3)$,
and because truncation deteriorates the preconditioning efficiency.
Finally, we reduce the time spent in the iterative preconditioning steps
by using a loose convergence threshold of $10^{-2}$ and by setting a
maximum number of iterations of 20. The convergence threshold
for initial guesses is set to $10^{-3}$.

\subsection{The impact of the maximum angular momentum}
\begin{figure}[tbp]
  \centering
  \includegraphics[width=\linewidth]{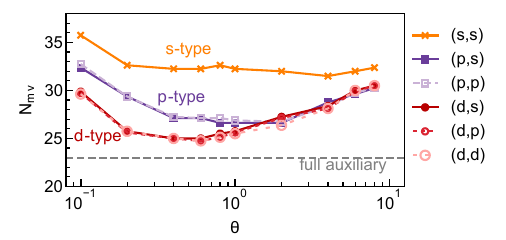}
      \caption{\label{fig:spd-theta}
      The average $N_\text{mv}$ obtained from computing lowest 5 states with
      PBE0/def2-SVP on TUNE8 set using different ris-type preconditioners.
      The label $(\ell_J,\ell_K)$ indicates a preconditioner using up
      to $\ell_J$ functions for the Coulomb terms and up to
      $\ell_K$ functions for the Exchange terms.
      For comparison, the dashed horizontal line shows
      the average $N_\text{mv}$ obtained
      using the full auxiliary basis as a preconditioner.
      }
\end{figure}

In this section we test the performance of
a series of 6 ris-type preconditioners with maximum angular momentum
up to d functions on non-hydrogen atoms.
In addition, we test using different maximum angular momentum
values for Coulomb and for exchange bases.
The preconditioners are labeled as $(\ell_J,\ell_K)$, where $\ell_J$ and
$\ell_K$ are the maximum angular momenta used for the Coulomb and exchange
terms, respectively. For example, the $(p,d)$ preconditioner uses up to
$p$ functions for the Coulomb terms and up to $d$ functions for the exchange
terms.
The average $N_\text{iter}$ across TUNE8 is shown for all 6 preconditioners
as a function of $\theta$ in Fig. \ref{fig:spd-theta}.
As comparison, the average $N_\text{mv}$ using the diagonal preconditioner is 50.5.
From this figure, we first see that there is indeed a slight
dependence of the optimal $\theta$ on the maximum angular momentum.
The preconditioner with only $s$ functions is relatively insensitive
to the value of $\theta$ but has maximum performance with $\theta=4$.
By contrast, the optimal $\theta$ found
by tuning the energetics and spectra was 0.2.\cite{zhou2023JPCL}
Preconditioners with up to $p$ functions, on the other hand, have an
optimal $\theta$ of 2.0, while preconditioners with up to $d$ functions
have an optimal $\theta$ of 0.6.
Next, we see that the performance of the preconditioners tends to improve
with the addition of angular momentum. The optimal $s$-only preconditioner has
an average $N_\text{mv}$ of 31.5, while the best $p$-type and $d$-type
preconditioners reduce this to 26.6 and 24.8, respectively.
Finally, we find, somewhat fortuitously, that the preconditioning
performance is essentially ambivalent about the angular momentum used
for the exchange terms. For example, the optimal $(p,s)$ type and $(p,p)$ type
preconditioners use $N_\text{mv}$ of 26.6 and 26.8, respectively.
Similarly, the optimal $(d,s)$, $(d,p)$, and $(d,d)$ preconditioners
have $N_\text{mv}$ of 25.0, 24.8, and 24.8, essentially identical
performance.

We do not consider higher angular momentum functions because the
$d$-type preconditioners are already nearly optimal. To demonstrate this,
we used the full auxiliary basis as a preconditioner. The full auxiliary basis
would be an impractical preconditioner, but it can show the upper limit
of what can be achieved with a preconditioner that neglects $f^\text{xc}$.
With the full auxiliary basis, the
test suite converges with $N_\text{mv}=23$ (see Fig. \ref{fig:spd-theta}),
only reducing the $(d,s)$ result by 2 matrix-vector products or 8\%.

Based on these results, we propose the $(d,s)$ preconditioner with
$\theta=0.6$ as our best performing preconditioner, and name it rid.
With this structure, the total cost of preconditioning is much less
than 1\% of the overall algorithm.
As a last step for creating the rid preconditioner, we re-tune
the number of initial guesses used to initiate the Davidson algorithm.
We find that the performance of the rid preconditioner is
maximized when using up to 3 additional initial guesses.
That is, we generate
$N_\text{init} = N_\text{states} + \text{min}(N_\text{states}, 3)$
initial guesses for the Davidson algorithm.
The
diagonal and sTDA preconditioners, by contrast, are most effective
when using 8 additional initial guesses. The detailed results of this
tuning are shown in Fig. S1 in Supporting Information.

\section{Evaluating the Rid Preconditioner}
\label{sec:results}

In this section, we benchmark the performance of the rid preconditioner in
computing excitation energies and polarizabilities for small to medium sized
molecules. We compare its efficacy to that of the diagonal preconditioner and
the sTDA preconditioner using the PBE0\cite{perdew1996JCP} and
$\omega$B97X\cite{Chai2008TJCP} density functionals with the def2-TZVP basis
set.\cite{weigend2005PCCP}
For consistency, we employ the same set of 19 molecules
from our previous study, referenced as PRECOND19.\cite{zhou2021JCP} This set
comprises nanoparticle, organic dyes, chemical probes, and bioluminescent
molecules (see the Supporting Information for the complete list of
molecules in PRECOND19). For clarity, TUNE8 is a subset of PRECOND19.
We use as figures of merit the number of iterations,
$N_\text{iter}$, as well as the number of ab initio matrix-vector products, $N_\text{mv}$.
The speedup factor, $\zeta_\text{iter}$ or $\zeta_\text{mv}$,
is defined as the ratio of $N_\text{iter}$ or $N_\text{mv}$ for the diagonal
preconditioner to that for the rid preconditioner. For example,
$\zeta^\text{rid}_\text{iter} = N^\text{diag}_\text{iter} / N^\text{rid}_\text{iter}$.
For tables with all the results shown in this section, see the Supporting
Information.

\begin{figure}[tbp]
  \centering
  \includegraphics[width=\linewidth]{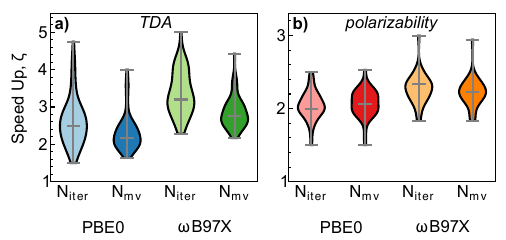}
      \caption{\label{fig:violin}
      Violin plot of all observed speedups using the rid preconditioner
      for computing a) TDA excitation energies and b) polarizabilities.
      For each dataset, the median, minimum, and maximum observations
      are marked with a line.
      }
\end{figure}

The complete results of these benchmarks are summarized in Fig. \ref{fig:violin},
which shows a violin plot of all the observed speedups using the rid preconditioner
relative to the diagonal preconditioner for both excitation energies and polarizabilities.

\subsection{TDA Excitation Energies}
\begin{figure}[tbp]
    \centering
    \includegraphics[width=\linewidth]{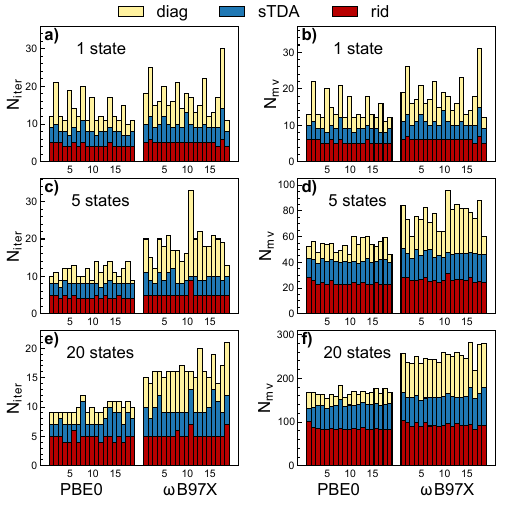}
    \caption{\label{fig:TDA_data}
        Performance of the diagonal, sTDA, and rid preconditioners
        for computing TDA excitation energies.
        Left three panels show the number of iterations ($N_\text{iter}$),
        while the right three panels show the matrix-vector products ($N_\text{mv}$)
        required to converge 1, 5 or 20 states.}
\end{figure}

To benchmark excitation energies, we compute the lowest 1, 5, and 20 states,
which are intended to mimic typical use cases of computing the optical gap,
low-lying states, and a UV-vis spectrum, respectively. The complete results
are shown in Fig. \ref{fig:TDA_data} and summarized in Table \ref{tab:TDA_N_iter}.
For more detailed results, see the Supporting Information.
Results using the full TDDFT equations (i.e., without the TDA) are similar
and are collected in Fig. S2 of the Supporting Information.

Fig. \ref{fig:TDA_data} shows that the rid preconditioner
systematically outperforms both the diagonal and the sTDA
preconditioners. For every molecule considered here, the rid preconditioner
converges faster than the sTDA preconditioner and the diagonal preconditioner.
Using PBE0, all excitation energies converge in at most 6
iterations, and only 4.5 iterations on average. By contrast,
all excitation energies required at least 7 iterations
using the sTDA preconditioner (8.0 on average) and at
least 9 iterations for the diagonal
preconditioner (11.6 on average), meaning the worst-case performance of the rid
preconditioner is better than the best-case performance of the conventional
and sTDA preconditioners. The rid preconditioner reduces the number of
iterations by a factor of 2.6 on average and up to a factor of 4.8, compared to
a factor of 1.5 on average and 2.7 at best for the sTDA.

The results are similar using $\omega$B97X: the rid preconditioner
converges in 5.2 iterations on average, compared to 9.8 for sTDA and 17.4
for the diagonal preconditioner.
Thus, the rid preconditioner reduces the number of iterations by a factor of
3.4 on average and up to a factor of 5.0, compared to a factor of 1.8 on average
and 3.3 at best for the sTDA.
Moreover, the rid preconditioner all but erases the
difference in the number of iterations required to converge
$\omega$B97X compared to PBE0. For example, using the diagonal
preconditioner, $\omega$B97X requires an additional 5.8 iterations than PBE0
on average, whereas with the rid preconditioner, only an additional 0.7 iterations
is required.

\begin{table}[tbp]
  \centering
  \caption{\label{tab:TDA_N_iter}
  The range and the average number of iterations, $N_\text{iter}$,
  required to compute TDA excitation energies using the PBE0 and $\omega$B97X
  density functionals with diagonal, sTDA and rid preconditioners.}
  \setlength{\tabcolsep}{1.0mm}{
    \begin{tabular}{cccccccc}
    \midrule\midrule
    & & \multicolumn{3}{c}{PBE0}& \multicolumn{3}{c}{$\omega$B97X}\\
      \cmidrule(lr){3-5} \cmidrule(lr){6-8}
    \multicolumn{2}{c}{precond.} & diag  & sTDA & rid  & diag  & sTDA & rid  \\
    \midrule
    \multirow{3}*{$N_\text{iter}$} & min. &	9	&	7	&	4 & 11 & 8 & 4 	\\
      & max.	&	21	&	11	&	6 & 33 & 14 & 9	\\
                                      & avg.	&	11.6	&	8.0	&	4.5 & 17.4 & 9.8 &5.2	\\
    \midrule
    \multirow{3}*{$\zeta_\text{iter}$}	&	min.	  & -- &	1.1	&	1.5	&--&	1.2	&	2.3	\\
        &	max.	&--&	2.7	&	4.8	   &--&	3.3	&	5.0	\\
                                         &	avg.	&--&	1.5	&	2.6	   &--&	1.8	&	3.4	\\
    \midrule\midrule
    \end{tabular}
    }
\end{table}

\begin{figure}[tbp]
  \centering
  \includegraphics[width=\linewidth]{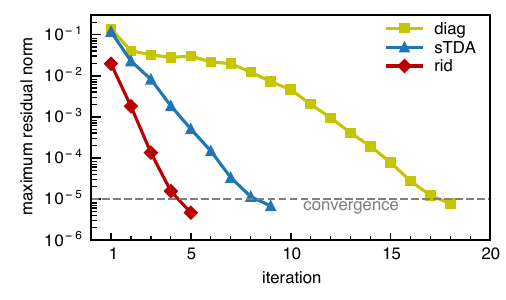}
      \caption{\label{fig:fluorescein_TDA_res}
      The convergence behavior of computing the lowest 5
      TDA excitation energies of fluorescein using $\omega$B97X/def2-TZVP
      with the diagonal, sTDA, and rid preconditioners.
      The horizontal line at
      $10^{-5}$ shows the convergence threshold.
      }
\end{figure}

For a more detailed analysis of the convergence behavior, we consider the
calculation of the first 5 excitation energies of fluorescein
(\textbf{5} in PRECOND19) using the $\omega$B97X functional.
The maximum residual norm at each iteration of the Davidson algorithm is shown
in Fig. \ref{fig:fluorescein_TDA_res}.
For this example, the rid preconditioner requires 5 iterations,
the sTDA preconditioner requires 9 iterations, and the diagonal 18.
From Fig. \ref{fig:fluorescein_TDA_res}, we see that compared to
the diagonal preconditioner, the rid algorithm is benefiting from
both an improved initial guess and from a faster convergence rate.
Consider the initial guesses first.
In Fig. \ref{fig:fluorescein_TDA_res}, we see that the diagonal preconditioner
stagnates for the first
8 iterations before the residual norm begins to decrease exponentially (note the
log scale), indicating that the algorithm is still searching for a suitable
starting vector. By contrast, the sTDA and rid initial guesses show no such
stagnation, because they immediately find suitable initial guesses.
Furthermore, the rid initial guess has a sharply reduced initial residual norm (0.020)
compared to sTDA (0.118) and the diagonal (0.139).

Next, consider the convergence rate.
Each algorithm in Fig. \ref{fig:fluorescein_TDA_res} converges essentially
exponentially once the residual norm falls below $10^{-2}$. We use this to estimate
rates of convergence for each algorithm by fitting a line to the base 10 logarithm of
the residual norm while it is in the range $10^{-2}$ to $10^{-5}$.
The rid preconditioner has a remarkable slope of $-1.03$, meaning that
the residual norm decreases by a factor of $10^{-1.03}=0.092$ in every iteration.
By contrast, the sTDA preconditioner has a slope of $-0.57$ ($10^{-0.57}=0.27$), and the
diagonal preconditioner has a slope of $-0.35$ ($10^{-0.35} = 0.44$).

While the improvement of sTDA over the diagonal preconditioner is nearly equal parts
due to the initial guess and the preconditioner, we find that the improvement of the
rid preconditioner over sTDA is almost entirely due to the preconditioner. The
second iteration of sTDA has a residual norm below the rid initial guess, meaning
that the sTDA initial guess ``catches up'' to the rid initial guess in one step.
However, the rid preconditioner still converges in 4 fewer iterations than sTDA.
This means that the rid preconditioner will retain its advantage over both
the sTDA and the diagonal preconditioners even if better initial guesses are used,
for example by using solutions from previous geometries in an optimization or in dynamics.
Furthermore, the advantage of the rid preconditioner will grow as the convergence threshold
is tightened, as the rid preconditioner has a faster convergence rate.

Finally, these fluorescein calculations illustrate the low cost of the preconditioning
step. The calculations using the rid preconditioner in Fig. \ref{fig:fluorescein_TDA_res}
required a total of 850 seconds of wall time on 16 CPUs, of which 11 seconds or about
1.2\% were spent in the preconditioning step \emph{even though the preconditioning
step is not yet parallelized}.

\subsection{Polarizability}
\begin{figure}[tbp]
  \centering
  \includegraphics[width=\linewidth]{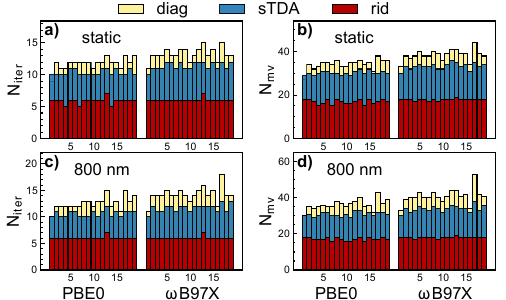}
      \caption{\label{fig:polar_data}
      Performance of semiempirical preconditioning for computing static
      and dynamic (800 nm) dipole polarizabilities. The top row shows
      a) $N_\text{iter}$ and b) $N_\text{mv}$ for static polarizabilities,
      and the bottom row shows c) $N_\text{iter}$ and d) $N_\text{mv}$
      for dynamic polarizabilities.
       }
\end{figure}

To benchmark polarizability calculations, we compute the static and
dynamic (using 800 nm light) polarizability tensors. The complete results
shown in Fig. \ref{fig:polar_data} and Table
\ref{tab:polar_N_iter_speedup}.

Similar to the excitation energy case, Fig. \ref{fig:polar_data} shows that
the rid preconditioner systematically improves over both the diagonal
and the sTDA preconditioners, with similar behavior for both the static and
dynamic polarizabilities. For PBE0, the rid preconditioner converges in
6.0 iterations on average, compared to 12.2 for the diagonal and 10.4 for the
sTDA. This leads to speed ups of a factor of 2.0 and 1.2 for rid and sTDA,
respectively. For $\omega$B97X, the rid preconditioner converges in 6.1
iterations on average, compared to 13.7 for the diagonal and 11.6 for the sTDA.
This gives speed ups of a factor of 2.3 and 1.2 for rid and sTDA,
respectively. Again, the rid preconditioner erases the difference in
number of iterations needed between PBE0 and $\omega$B97X.
We emphasize that the rid preconditioner fixes the disappointing performance
of the sTDA preconditioner for the polarizability problem.

\begin{table}[tbp]
  \centering
  \caption{\label{tab:polar_N_iter_speedup}
  The range and the average number of iterations, $N_\text{iter}$, required
  to converge static and dynamic (800 nm) polarizability calculations
  with the PBE0 and $\omega$B97X density functionals
  using diagonal, sTDA, and rid preconditioners.}
  \setlength{\tabcolsep}{1.0mm}{
    \begin{tabular}{cccccccc}
    \midrule\midrule
     & & \multicolumn{3}{c}{PBE0}& \multicolumn{3}{c}{$\omega$B97X}\\
      \cmidrule(lr){3-5} \cmidrule(lr){6-8}
    \multicolumn{2}{c}{precond.} & diag  & sTDA & rid  & diag  & sTDA & rid  \\
    \midrule
    \multirow{3}*{$N_\text{iter}$}    & min.	  &	 9 & 10 & 5 & 11 & 10 & 6 \\
      & max.	   & 15 & 12 & 7  & 18 & 13 & 7	\\
                                      & avg.	   & 12.2 & 10.4& 6.0  & 13.7 & 11.6& 6.1	\\
    \midrule
    \multirow{3}*{$\zeta_\text{iter}$} &	min.  & -- &	0.9 & 1.5	& -- & 1.1 & 1.8	\\
      &	max.   & -- &	1.4 & 2.5           & -- & 1.4 & 3.0	\\
                                       &	avg.   & -- &	1.2 & 2.0           & -- & 1.2 & 2.3	\\
    \midrule\midrule
    \end{tabular}
    }
\end{table}

\section{Conclusions}
\label{sec:conclusion}

TDDFT excitation energy and polarizability calculations are computationally
expensive for large systems. Semiempirical models, like the TDDFT-ris model,
are often designed to replace the need for ab initio calculations.
However, we show that semiempirical models and ab initio models can be
synergistically combined to accelerate ab initio calculations without
any loss in accuracy.

We used the TDDFT-ris model as a starting point to systematically design a
highly performant
and broadly applicable preconditioner for calculations of excitation energies
and polarizabilities within TDDFT. The final preconditioner, rid,
uses an auxiliary basis with up to $d$ functions for the Coulomb terms
and $s$ functions for the exchange terms, and a global scale factor of
$\theta$=0.6. By construction, the rid preconditioner has
virtually no additional cost compared to the ab initio matrix-vector
products, and is applicable across the entire periodic table.

The rid preconditioner significantly outperforms both the diagonal and
sTDA preconditioners for excitation energy and polarizability calculations.
We find that the rid preconditioner converges in just 5-6 iterations on average
for excitation energies and polarizabilities.
In other words, the rid preconditioner speeds up excitation energy calculations
by a factor of 2-5 compared to the diagonal preconditioner and
speeds up polarizability calculations by a factor of 1.3-2.2.
This makes the rid preconditioner the best performing preconditioner
for TDDFT excitation energies and polarizabilities that we are aware of.

We emphasize two further points about the rid preconditioner.
First, the rid preconditioner all but erases the difference in the
number of iterations required to converge PBE0 compared to $\omega$B97X,
indicating that it retains high performance across different density
functionals. Second, the rid preconditioner successfully speeds up
the polarizability calculations, thus fixing the disappointing performance
of the sTDA preconditioner for the polarizability problem.

The excellent performance of the rid preconditioner may appear surprising
in light of the common view that the Davidson algorithm can
stagnate if the approximate matrix used in the preconditioner, $\mathbf{T}$,
becomes ``too close'' to the exact matrix,
$\mathbf{A}$.\cite{Olsen1990CPL,Sleijpen1996SJMAA,Windom2023JCP}
However, this ``deficiency'' in the Davidson algorithm turns out have little
relevance to practical applications in quantum chemistry. As shown by Notay, when
the Davidson algorithm is initiated with eigenvectors of $\mathbf{T}$---as we
do here and as is common practice---then
no stagnation occurs and both Davidson and Jacobi-Davidson provide similar
convergence rates.\cite{Notay2004SJMAA}
Our results support this conclusion, as we find no evidence at all of
any deterioration in the performance of the Davidson algorithm when using
increasingly accurate preconditioners.

Thus, the rid preconditioner is a general purpose
preconditioner that can be used to accelerate the iterative calculation
of TDDFT excitation energies and polarizabilities.
In addition, because rid is fundamentally based on a transferrable
approximation to the electron repulsion integrals, the same idea could be used
to design preconditioners in other contexts. A rid-type preconditioner
should be beneficial for any method where computing electron repulsion integrals
are the bottleneck. We are especially eager to use the rid preconditioner
in scenarios where the electronic Hessian are intensively
called, such as non-adiabatic molecular dynamics, and excited state geometry
optimization. We also envision the rid preconditioner proving invaluable
when applied to vibrational frequency calculations,
and in generating large datasets for machine learning models.

\begin{suppinfo}
See the Supporting Information for definitions of the TUNE8 and PRECOND19
benchmark sets, results from tuning the number of initial guesses,
the performance of the rid preconditioner for TDDFT eigenvalues,
and the detailed set of results for all benchmark calculations.
\end{suppinfo}

\begin{acknowledgement}
This work was supported by a startup fund from Case Western Reserve University.
This work made use of the High Performance Computing Resource in the Core
Facility for Advanced Research Computing at Case Western Reserve University.
\end{acknowledgement}

\section*{Data Availability}
All code for this paper is available on Github under the MIT
license.\cite{davidsoncode}
The data supporting this study are openly available on the Open Science Framework
under the Creative Commons
license.\cite{osfdata}

\bibliography{ris_precond}

\providecommand{\latin}[1]{#1}
\makeatletter
\providecommand{\doi}
  {\begingroup\let\do\@makeother\dospecials
  \catcode`\{=1 \catcode`\}=2 \doi@aux}
\providecommand{\doi@aux}[1]{\endgroup\texttt{#1}}
\makeatother
\providecommand*\mcitethebibliography{\thebibliography}
\csname @ifundefined\endcsname{endmcitethebibliography}
  {\let\endmcitethebibliography\endthebibliography}{}
\begin{mcitethebibliography}{48}
\providecommand*\natexlab[1]{#1}
\providecommand*\mciteSetBstSublistMode[1]{}
\providecommand*\mciteSetBstMaxWidthForm[2]{}
\providecommand*\mciteBstWouldAddEndPuncttrue
  {\def\EndOfBibitem{\unskip.}}
\providecommand*\mciteBstWouldAddEndPunctfalse
  {\let\EndOfBibitem\relax}
\providecommand*\mciteSetBstMidEndSepPunct[3]{}
\providecommand*\mciteSetBstSublistLabelBeginEnd[3]{}
\providecommand*\EndOfBibitem{}
\mciteSetBstSublistMode{f}
\mciteSetBstMaxWidthForm{subitem}{(\alph{mcitesubitemcount})}
\mciteSetBstSublistLabelBeginEnd
  {\mcitemaxwidthsubitemform\space}
  {\relax}
  {\relax}

\bibitem[Siegbahn(1977)]{Siegbahn1977CP}
Siegbahn,~P. E.~M. The Direct Configuration Interaction Method with a
  Contracted Configuration Expansion. \emph{Chem. Phys.} \textbf{1977},
  \emph{25}, 197--205\relax
\mciteBstWouldAddEndPuncttrue
\mciteSetBstMidEndSepPunct{\mcitedefaultmidpunct}
{\mcitedefaultendpunct}{\mcitedefaultseppunct}\relax
\EndOfBibitem
\bibitem[Knowles and Handy(1984)Knowles, and Handy]{Knowles1984CPL}
Knowles,~P.~J.; Handy,~N.~C. A New Determinant-Based Full Configuration
  Interaction Method. \emph{Chem. Phys. Lett.} \textbf{1984}, \emph{111},
  315--321\relax
\mciteBstWouldAddEndPuncttrue
\mciteSetBstMidEndSepPunct{\mcitedefaultmidpunct}
{\mcitedefaultendpunct}{\mcitedefaultseppunct}\relax
\EndOfBibitem
\bibitem[Olsen \latin{et~al.}(1990)Olsen, J{\o}rgensen, and
  Simons]{Olsen1990CPL}
Olsen,~J.; J{\o}rgensen,~P.; Simons,~J. Passing the One-Billion Limit in Full
  Configuration-Interaction ({{FCI}}) Calculations. \emph{Chem. Phys. Lett.}
  \textbf{1990}, \emph{169}, 463--472\relax
\mciteBstWouldAddEndPuncttrue
\mciteSetBstMidEndSepPunct{\mcitedefaultmidpunct}
{\mcitedefaultendpunct}{\mcitedefaultseppunct}\relax
\EndOfBibitem
\bibitem[Mitrushenkov(1994)]{Mitrushenkov1994CPL}
Mitrushenkov,~A.~O. Passing the Several Billions Limit in {{FCI}} Calculations
  on a Mini-Computer. \emph{Chem. Phys. Lett.} \textbf{1994}, \emph{217},
  559--565\relax
\mciteBstWouldAddEndPuncttrue
\mciteSetBstMidEndSepPunct{\mcitedefaultmidpunct}
{\mcitedefaultendpunct}{\mcitedefaultseppunct}\relax
\EndOfBibitem
\bibitem[Gao \latin{et~al.}(2024)Gao, Imamura, Kasagi, and
  Yoshida]{Gao2024JCTC}
Gao,~H.; Imamura,~S.; Kasagi,~A.; Yoshida,~E. Distributed {{Implementation}} of
  {{Full Configuration Interaction}} for {{One Trillion Determinants}}.
  \emph{J. Chem. Theory Comput.} \textbf{2024}, \emph{20}, 1185--1192\relax
\mciteBstWouldAddEndPuncttrue
\mciteSetBstMidEndSepPunct{\mcitedefaultmidpunct}
{\mcitedefaultendpunct}{\mcitedefaultseppunct}\relax
\EndOfBibitem
\bibitem[Olsen and J{\o}rgensen(1985)Olsen, and J{\o}rgensen]{Olsen1985JCP}
Olsen,~J.; J{\o}rgensen,~P. Linear and Nonlinear Response Functions for an
  Exact State and for an {{MCSCF}} State. \emph{J. Chem. Phys.} \textbf{1985},
  \emph{82}, 3235--3264\relax
\mciteBstWouldAddEndPuncttrue
\mciteSetBstMidEndSepPunct{\mcitedefaultmidpunct}
{\mcitedefaultendpunct}{\mcitedefaultseppunct}\relax
\EndOfBibitem
\bibitem[Christiansen \latin{et~al.}(1998)Christiansen, J{\o}rgensen, and
  H{\"a}ttig]{Christiansen1998IJQC}
Christiansen,~O.; J{\o}rgensen,~P.; H{\"a}ttig,~C. Response Functions from
  {{Fourier}} Component Variational Perturbation Theory Applied to a
  Time-Averaged Quasienergy. \emph{Int. J. Quantum Chem.} \textbf{1998}, \relax
\mciteBstWouldAddEndPunctfalse
\mciteSetBstMidEndSepPunct{\mcitedefaultmidpunct}
{}{\mcitedefaultseppunct}\relax
\EndOfBibitem
\bibitem[Parker and Furche(2018)Parker, and Furche]{Parker2018FoQC}
Parker,~S.~M.; Furche,~F. In \emph{Frontiers of {{Quantum Chemistry}}};
  W{\'o}jcik,~M.~J., Nakatsuji,~H., Kirtman,~B., Ozaki,~Y., Eds.; Springer
  Singapore, 2018; pp 69--86\relax
\mciteBstWouldAddEndPuncttrue
\mciteSetBstMidEndSepPunct{\mcitedefaultmidpunct}
{\mcitedefaultendpunct}{\mcitedefaultseppunct}\relax
\EndOfBibitem
\bibitem[Thouless(1960)]{Thouless1960NP}
Thouless,~D.~J. Stability Conditions and Nuclear Rotations in the
  {{Hartree-Fock}} Theory. \emph{Nuc. Phys.} \textbf{1960}, \emph{21},
  225--232\relax
\mciteBstWouldAddEndPuncttrue
\mciteSetBstMidEndSepPunct{\mcitedefaultmidpunct}
{\mcitedefaultendpunct}{\mcitedefaultseppunct}\relax
\EndOfBibitem
\bibitem[{\v C}{\'i}{\v z}ek and Paldus(1967){\v C}{\'i}{\v z}ek, and
  Paldus]{Cizek1967JCP}
{\v C}{\'i}{\v z}ek,~J.; Paldus,~J. Stability {{Conditions}} for the
  {{Solutions}} of the {{Hartree}}---{{Fock Equations}} for {{Atomic}} and
  {{Molecular Systems}}. {{Application}} to the {{Pi}}-{{Electron Model}} of
  {{Cyclic Polyenes}}. \emph{J. Chem. Phys.} \textbf{1967}, \emph{47},
  3976--3985\relax
\mciteBstWouldAddEndPuncttrue
\mciteSetBstMidEndSepPunct{\mcitedefaultmidpunct}
{\mcitedefaultendpunct}{\mcitedefaultseppunct}\relax
\EndOfBibitem
\bibitem[Bauernschmitt and Ahlrichs(1996)Bauernschmitt, and
  Ahlrichs]{Bauernschmitt1996JCP}
Bauernschmitt,~R.; Ahlrichs,~R. Stability Analysis for Solutions of the Closed
  Shell {{Kohn}}--{{Sham}} Equation. \emph{J. Chem. Phys.} \textbf{1996},
  \emph{104}, 9047\relax
\mciteBstWouldAddEndPuncttrue
\mciteSetBstMidEndSepPunct{\mcitedefaultmidpunct}
{\mcitedefaultendpunct}{\mcitedefaultseppunct}\relax
\EndOfBibitem
\bibitem[Davidson(1975)]{davidson1975JCoP}
Davidson,~E.~R. The Iterative Calculation of a Few of the Lowest Eigenvalues
  and Corresponding Eigenvectors of Large Real-Symmetric Matrices. \emph{J.
  Comput. Phys.} \textbf{1975}, \emph{17}, 87--94\relax
\mciteBstWouldAddEndPuncttrue
\mciteSetBstMidEndSepPunct{\mcitedefaultmidpunct}
{\mcitedefaultendpunct}{\mcitedefaultseppunct}\relax
\EndOfBibitem
\bibitem[Weiss \latin{et~al.}(1993)Weiss, Ahlrichs, and
  H{\"a}ser]{Weiss1993JCP}
Weiss,~H.; Ahlrichs,~R.; H{\"a}ser,~M. A Direct Algorithm for
  Self-consistent-field Linear Response Theory and Application to {{C60}}:
  {{Excitation}} Energies, Oscillator Strengths, and Frequency-dependent
  Polarizabilities. \emph{J. Chem. Phys.} \textbf{1993}, \emph{99},
  1262--1270\relax
\mciteBstWouldAddEndPuncttrue
\mciteSetBstMidEndSepPunct{\mcitedefaultmidpunct}
{\mcitedefaultendpunct}{\mcitedefaultseppunct}\relax
\EndOfBibitem
\bibitem[Furche \latin{et~al.}(2016)Furche, Krull, Nguyen, and
  Kwon]{furche2016JCP}
Furche,~F.; Krull,~B.~T.; Nguyen,~B.~D.; Kwon,~J. Accelerating Molecular
  Property Calculations with Nonorthonormal Krylov Space Methods. \emph{J.
  Chem. Phys.} \textbf{2016}, \emph{144}, 174105\relax
\mciteBstWouldAddEndPuncttrue
\mciteSetBstMidEndSepPunct{\mcitedefaultmidpunct}
{\mcitedefaultendpunct}{\mcitedefaultseppunct}\relax
\EndOfBibitem
\bibitem[Parrish \latin{et~al.}(2016)Parrish, Hohenstein, and
  Mart{\'i}nez]{Parrish2016JCTC}
Parrish,~R.~M.; Hohenstein,~E.~G.; Mart{\'i}nez,~T.~J. ``{{Balancing}}'' the
  {{Block Davidson}}--{{Liu Algorithm}}. \emph{J. Chem. Theory Comput.}
  \textbf{2016}, \emph{12}, 3003--3007\relax
\mciteBstWouldAddEndPuncttrue
\mciteSetBstMidEndSepPunct{\mcitedefaultmidpunct}
{\mcitedefaultendpunct}{\mcitedefaultseppunct}\relax
\EndOfBibitem
\bibitem[Bauernschmitt \latin{et~al.}(1997)Bauernschmitt, H{\"a}ser, Treutler,
  and Ahlrichs]{Bauernschmitt1997CPL}
Bauernschmitt,~R.; H{\"a}ser,~M.; Treutler,~O.; Ahlrichs,~R. Calculation of
  Excitation Energies within Time-Dependent Density Functional Theory Using
  Auxiliary Basis Set Expansions. \emph{Chem. Phys. Lett.} \textbf{1997},
  \emph{264}, 573--578\relax
\mciteBstWouldAddEndPuncttrue
\mciteSetBstMidEndSepPunct{\mcitedefaultmidpunct}
{\mcitedefaultendpunct}{\mcitedefaultseppunct}\relax
\EndOfBibitem
\bibitem[Hu \latin{et~al.}(2020)Hu, Liu, Li, Ding, Yang, and Yang]{Hu2020JCTC}
Hu,~W.; Liu,~J.; Li,~Y.; Ding,~Z.; Yang,~C.; Yang,~J. Accelerating {{Excitation
  Energy Computation}} in {{Molecules}} and {{Solids}} within {{Linear-Response
  Time-Dependent Density Functional Theory}} via {{Interpolative Separable
  Density Fitting Decomposition}}. \emph{J. Chem. Theory Comput.}
  \textbf{2020}, \emph{16}, 964--973\relax
\mciteBstWouldAddEndPuncttrue
\mciteSetBstMidEndSepPunct{\mcitedefaultmidpunct}
{\mcitedefaultendpunct}{\mcitedefaultseppunct}\relax
\EndOfBibitem
\bibitem[Ufimtsev and Mart{\'i}nez(2008)Ufimtsev, and
  Mart{\'i}nez]{Ufimtsev2008JCTC}
Ufimtsev,~I.~S.; Mart{\'i}nez,~T.~J. Quantum {{Chemistry}} on {{Graphical
  Processing Units}}. 1. {{Strategies}} for {{Two-Electron Integral
  Evaluation}}. \emph{J. Chem. Theory Comput.} \textbf{2008}, \emph{4},
  222--231\relax
\mciteBstWouldAddEndPuncttrue
\mciteSetBstMidEndSepPunct{\mcitedefaultmidpunct}
{\mcitedefaultendpunct}{\mcitedefaultseppunct}\relax
\EndOfBibitem
\bibitem[Isborn \latin{et~al.}(2011)Isborn, Luehr, Ufimtsev, and
  Mart{\'i}nez]{Isborn2011JCTC}
Isborn,~C.~M.; Luehr,~N.; Ufimtsev,~I.~S.; Mart{\'i}nez,~T.~J. Excited-{{State
  Electronic Structure}} with {{Configuration Interaction Singles}} and
  {{Tamm}}--{{Dancoff Time-Dependent Density Functional Theory}} on {{Graphical
  Processing Units}}. \emph{J. Chem. Theory Comput.} \textbf{2011}, \emph{7},
  1814--1823\relax
\mciteBstWouldAddEndPuncttrue
\mciteSetBstMidEndSepPunct{\mcitedefaultmidpunct}
{\mcitedefaultendpunct}{\mcitedefaultseppunct}\relax
\EndOfBibitem
\bibitem[Zhou and Parker(2021)Zhou, and Parker]{zhou2021JCP}
Zhou,~Z.; Parker,~S.~M. Accelerating molecular property calculations with
  semiempirical preconditioning. \emph{J. Chem. Phys.} \textbf{2021},
  \emph{155}, 204111\relax
\mciteBstWouldAddEndPuncttrue
\mciteSetBstMidEndSepPunct{\mcitedefaultmidpunct}
{\mcitedefaultendpunct}{\mcitedefaultseppunct}\relax
\EndOfBibitem
\bibitem[Grimme(2013)]{grimme2013JCP}
Grimme,~S. A Simplified Tamm-Dancoff Density Functional Approach for the
  Electronic Excitation Spectra of Very Large Molecules. \emph{J. Chem. Phys.}
  \textbf{2013}, \emph{138}, 244104\relax
\mciteBstWouldAddEndPuncttrue
\mciteSetBstMidEndSepPunct{\mcitedefaultmidpunct}
{\mcitedefaultendpunct}{\mcitedefaultseppunct}\relax
\EndOfBibitem
\bibitem[Zhou \latin{et~al.}(2023)Zhou, Della~Sala, and Parker]{zhou2023JPCL}
Zhou,~Z.; Della~Sala,~F.; Parker,~S.~M. Minimal Auxiliary Basis Set Approach
  for the Electronic Excitation Spectra of Organic Molecules. \emph{J. Phys.
  Chem. Lett.} \textbf{2023}, 1968--1976\relax
\mciteBstWouldAddEndPuncttrue
\mciteSetBstMidEndSepPunct{\mcitedefaultmidpunct}
{\mcitedefaultendpunct}{\mcitedefaultseppunct}\relax
\EndOfBibitem
\bibitem[Sleijpen and {Van der Vorst}(1996)Sleijpen, and {Van der
  Vorst}]{Sleijpen1996SJMAA}
Sleijpen,~G.~L.; {Van der Vorst},~H.~A. A {{Jacobi}}--{{Davidson Iteration
  Method}} for {{Linear Eigenvalue Problems}}. \emph{SIAM Journal on Matrix
  Analysis and Applications} \textbf{1996}, \emph{17}, 401--425\relax
\mciteBstWouldAddEndPuncttrue
\mciteSetBstMidEndSepPunct{\mcitedefaultmidpunct}
{\mcitedefaultendpunct}{\mcitedefaultseppunct}\relax
\EndOfBibitem
\bibitem[Hochstenbach and Notay(2006)Hochstenbach, and
  Notay]{Hochstenbach2006G}
Hochstenbach,~M.; Notay,~Y. The {{Jacobi}}--{{Davidson}} Method.
  \emph{GAMM-Mitteilungen} \textbf{2006}, \emph{29}, 368--382\relax
\mciteBstWouldAddEndPuncttrue
\mciteSetBstMidEndSepPunct{\mcitedefaultmidpunct}
{\mcitedefaultendpunct}{\mcitedefaultseppunct}\relax
\EndOfBibitem
\bibitem[Van~Dam \latin{et~al.}(1996)Van~Dam, Van~Lenthe, Sleijpen, and Van
  Der~Vorst]{VanDam1996JCC}
Van~Dam,~H.; Van~Lenthe,~J.; Sleijpen,~G.; Van Der~Vorst,~H. An Improvement of
  {{Davidson}}'s Iteration Method: {{Applications}} to {{MRCI}} and {{MRCEPA}}
  Calculations. \emph{J. Comp. Chem.} \textbf{1996}, \emph{17}, 267--272\relax
\mciteBstWouldAddEndPuncttrue
\mciteSetBstMidEndSepPunct{\mcitedefaultmidpunct}
{\mcitedefaultendpunct}{\mcitedefaultseppunct}\relax
\EndOfBibitem
\bibitem[Rappoport \latin{et~al.}(2023)Rappoport, Bekoe, Mohanam, Le, George,
  Shen, and Furche]{Rappoport2023JCC}
Rappoport,~D.; Bekoe,~S.; Mohanam,~L.~N.; Le,~S.; George,~N.; Shen,~Z.;
  Furche,~F. Libkrylov: {{A}} Modular Open-Source Software Library for
  Extremely Large on-the-Fly Matrix Computations. \emph{J. Comp. Chem.}
  \textbf{2023}, \emph{44}, 1105--1118\relax
\mciteBstWouldAddEndPuncttrue
\mciteSetBstMidEndSepPunct{\mcitedefaultmidpunct}
{\mcitedefaultendpunct}{\mcitedefaultseppunct}\relax
\EndOfBibitem
\bibitem[Notay(2004)]{Notay2004SJMAA}
Notay,~Y. Is {{Jacobi--Davidson Faster}} than {{Davidson}}? \emph{SIAM Journal
  on Matrix Analysis and Applications} \textbf{2004}, \emph{26}, 522--543\relax
\mciteBstWouldAddEndPuncttrue
\mciteSetBstMidEndSepPunct{\mcitedefaultmidpunct}
{\mcitedefaultendpunct}{\mcitedefaultseppunct}\relax
\EndOfBibitem
\bibitem[Bannwarth and Grimme(2014)Bannwarth, and Grimme]{bannwarth2014CTC}
Bannwarth,~C.; Grimme,~S. A Simplified Time-Dependent Density Functional Theory
  Approach for Electronic Ultraviolet and Circular Dichroism Spectra of Very
  Large Molecules. \emph{Comput. Theor. Chem.} \textbf{2014}, \emph{1040-1041},
  45--53\relax
\mciteBstWouldAddEndPuncttrue
\mciteSetBstMidEndSepPunct{\mcitedefaultmidpunct}
{\mcitedefaultendpunct}{\mcitedefaultseppunct}\relax
\EndOfBibitem
\bibitem[{Asadi-Aghbolaghi} \latin{et~al.}(2020){Asadi-Aghbolaghi}, R{\"u}ger,
  Jamshidi, and Visscher]{asadi-aghbolaghi2020JPCC}
{Asadi-Aghbolaghi},~N.; R{\"u}ger,~R.; Jamshidi,~Z.; Visscher,~L. TD-DFT+TB: An
  Efficient and Fast Approach for Quantum Plasmonic Excitations. \emph{J. Phys.
  Chem. C} \textbf{2020}, \emph{124}, 7946--7955\relax
\mciteBstWouldAddEndPuncttrue
\mciteSetBstMidEndSepPunct{\mcitedefaultmidpunct}
{\mcitedefaultendpunct}{\mcitedefaultseppunct}\relax
\EndOfBibitem
\bibitem[Giannone and Della~Sala(2020)Giannone, and
  Della~Sala]{giannone2020JCP}
Giannone,~G.; Della~Sala,~F. Minimal auxiliary basis set for time-dependent
  density functional theory and comparison with tight-binding approximations:
  Application to silver nanoparticles. \emph{J. Chem. Phys.} \textbf{2020},
  \emph{153}, 084110\relax
\mciteBstWouldAddEndPuncttrue
\mciteSetBstMidEndSepPunct{\mcitedefaultmidpunct}
{\mcitedefaultendpunct}{\mcitedefaultseppunct}\relax
\EndOfBibitem
\bibitem[Baerends \latin{et~al.}(1973)Baerends, Ellis, and Ros]{baerends1973CP}
Baerends,~E.; Ellis,~D.; Ros,~P. Self-consistent molecular Hartree\textemdash
  Fock\textemdash Slater calculations I. The computational procedure.
  \emph{Chem. Phys.} \textbf{1973}, \emph{2}, 41\relax
\mciteBstWouldAddEndPuncttrue
\mciteSetBstMidEndSepPunct{\mcitedefaultmidpunct}
{\mcitedefaultendpunct}{\mcitedefaultseppunct}\relax
\EndOfBibitem
\bibitem[Dunlap \latin{et~al.}(1979)Dunlap, Connolly, and Sabin]{dunlap1979JCP}
Dunlap,~B.~I.; Connolly,~J. W.~D.; Sabin,~J.~R. On Some Approximations in
  Applications of X$\alpha$ Theory. \emph{J. Chem. Phys.} \textbf{1979},
  \emph{71}, 3396--3402\relax
\mciteBstWouldAddEndPuncttrue
\mciteSetBstMidEndSepPunct{\mcitedefaultmidpunct}
{\mcitedefaultendpunct}{\mcitedefaultseppunct}\relax
\EndOfBibitem
\bibitem[Eichkorn \latin{et~al.}(1995)Eichkorn, Treutler, {\"O}hm, H{\"a}ser,
  and Ahlrichs]{eichkorn1995CPL}
Eichkorn,~K.; Treutler,~O.; {\"O}hm,~H.; H{\"a}ser,~M.; Ahlrichs,~R. Auxiliary
  basis sets to approximate Coulomb potentials. \emph{Chem. Phys. Lett.}
  \textbf{1995}, \emph{240}, 283\relax
\mciteBstWouldAddEndPuncttrue
\mciteSetBstMidEndSepPunct{\mcitedefaultmidpunct}
{\mcitedefaultendpunct}{\mcitedefaultseppunct}\relax
\EndOfBibitem
\bibitem[Heinze \latin{et~al.}(2000)Heinze, G{\"o}rling, and
  R{\"o}sch]{heinze2000JCP}
Heinze,~H.~H.; G{\"o}rling,~A.; R{\"o}sch,~N. An efficient method for
  calculating molecular excitation energies by time-dependent
  density-functional theory. \emph{J. Chem. Phys.} \textbf{2000}, \emph{113},
  2088\relax
\mciteBstWouldAddEndPuncttrue
\mciteSetBstMidEndSepPunct{\mcitedefaultmidpunct}
{\mcitedefaultendpunct}{\mcitedefaultseppunct}\relax
\EndOfBibitem
\bibitem[Neese and Olbrich(2002)Neese, and Olbrich]{neese2002CPL}
Neese,~F.; Olbrich,~G. Efficient use of the resolution of the identity
  approximation in time-dependent density functional calculations with hybrid
  density functionals. \emph{Chem. Phys. Lett.} \textbf{2002}, \emph{362},
  170--178\relax
\mciteBstWouldAddEndPuncttrue
\mciteSetBstMidEndSepPunct{\mcitedefaultmidpunct}
{\mcitedefaultendpunct}{\mcitedefaultseppunct}\relax
\EndOfBibitem
\bibitem[Pedersen \latin{et~al.}(2009)Pedersen, Aquilante, and
  Lindh]{pedersenthomasbondo2009TCA}
Pedersen,~T.; Aquilante,~F.; Lindh,~R. Density fitting with auxiliary basis
  sets from Cholesky decompositions. \emph{Theor. Chem. Acc.} \textbf{2009},
  \emph{124}, 1\relax
\mciteBstWouldAddEndPuncttrue
\mciteSetBstMidEndSepPunct{\mcitedefaultmidpunct}
{\mcitedefaultendpunct}{\mcitedefaultseppunct}\relax
\EndOfBibitem
\bibitem[Weigend \latin{et~al.}(2009)Weigend, Kattannek, and
  Ahlrichs]{weigend2009JCP}
Weigend,~F.; Kattannek,~M.; Ahlrichs,~R. Approximated electron repulsion
  integrals: Cholesky decomposition versus resolution of the identity methods.
  \emph{J. Chem. Phys.} \textbf{2009}, \emph{130}, 164106\relax
\mciteBstWouldAddEndPuncttrue
\mciteSetBstMidEndSepPunct{\mcitedefaultmidpunct}
{\mcitedefaultendpunct}{\mcitedefaultseppunct}\relax
\EndOfBibitem
\bibitem[Stoychev \latin{et~al.}(2017)Stoychev, Auer, and
  Neese]{stoychev2017JCTC}
Stoychev,~G.~L.; Auer,~A.~A.; Neese,~F. Automatic generation of auxiliary basis
  sets. \emph{J. Chem. Theory Comput.} \textbf{2017}, \emph{13}, 554\relax
\mciteBstWouldAddEndPuncttrue
\mciteSetBstMidEndSepPunct{\mcitedefaultmidpunct}
{\mcitedefaultendpunct}{\mcitedefaultseppunct}\relax
\EndOfBibitem
\bibitem[Ghosh \latin{et~al.}(2008)Ghosh, Biswas, Chakraborty, Islam, and
  Rajak]{ghosh2008JMS}
Ghosh,~D.~C.; Biswas,~R.; Chakraborty,~T.; Islam,~N.; Rajak,~S.~K. The wave
  mechanical evaluation of the absolute radii of atoms. \emph{Journal of
  Molecular Structure: THEOCHEM} \textbf{2008}, \emph{865}, 60--67\relax
\mciteBstWouldAddEndPuncttrue
\mciteSetBstMidEndSepPunct{\mcitedefaultmidpunct}
{\mcitedefaultendpunct}{\mcitedefaultseppunct}\relax
\EndOfBibitem
\bibitem[Risthaus \latin{et~al.}(2014)Risthaus, Hansen, and
  Grimme]{Risthaus2014PCCP}
Risthaus,~T.; Hansen,~A.; Grimme,~S. Excited States Using the Simplified
  {{Tamm}}{\textendash}{{Dancoff-Approach}} for Range-Separated Hybrid Density
  Functionals: Development and Application. \emph{Phys. Chem. Chem. Phys.}
  \textbf{2014}, \emph{16}, 14408--14419\relax
\mciteBstWouldAddEndPuncttrue
\mciteSetBstMidEndSepPunct{\mcitedefaultmidpunct}
{\mcitedefaultendpunct}{\mcitedefaultseppunct}\relax
\EndOfBibitem
\bibitem[Sun \latin{et~al.}(2018)Sun, Berkelbach, Blunt, Booth, Guo, Li, Liu,
  McClain, Sayfutyarova, Sharma, Wouters, and Chan]{sun2018WCMS}
Sun,~Q.; Berkelbach,~T.~C.; Blunt,~N.~S.; Booth,~G.~H.; Guo,~S.; Li,~Z.;
  Liu,~J.; McClain,~J.~D.; Sayfutyarova,~E.~R.; Sharma,~S.; Wouters,~S.;
  Chan,~G. K.-L. PySCF: The Python-Based Simulations of Chemistry Framework.
  \emph{WIREs Comput. Mol. Sci.} \textbf{2018}, \emph{8}, e1340\relax
\mciteBstWouldAddEndPuncttrue
\mciteSetBstMidEndSepPunct{\mcitedefaultmidpunct}
{\mcitedefaultendpunct}{\mcitedefaultseppunct}\relax
\EndOfBibitem
\bibitem[Perdew \latin{et~al.}(1996)Perdew, Ernzerhof, and
  Burke]{perdew1996JCP}
Perdew,~J.~P.; Ernzerhof,~M.; Burke,~K. Rationale for Mixing Exact Exchange
  with Density Functional Approximations. \emph{J. Chem. Phys.} \textbf{1996},
  \emph{105}, 9982--9985\relax
\mciteBstWouldAddEndPuncttrue
\mciteSetBstMidEndSepPunct{\mcitedefaultmidpunct}
{\mcitedefaultendpunct}{\mcitedefaultseppunct}\relax
\EndOfBibitem
\bibitem[Weigend and Ahlrichs(2005)Weigend, and Ahlrichs]{weigend2005PCCP}
Weigend,~F.; Ahlrichs,~R. Balanced Basis Sets of Split Valence, Triple Zeta
  Valence and Quadruple Zeta Valence Quality for H to Rn: Design and Assessment
  of Accuracy. \emph{Phys. Chem. Chem. Phys.} \textbf{2005}, \emph{7},
  3297--3305\relax
\mciteBstWouldAddEndPuncttrue
\mciteSetBstMidEndSepPunct{\mcitedefaultmidpunct}
{\mcitedefaultendpunct}{\mcitedefaultseppunct}\relax
\EndOfBibitem
\bibitem[Chai and {Head-Gordon}(2008)Chai, and {Head-Gordon}]{Chai2008TJCP}
Chai,~J.-D.; {Head-Gordon},~M. Systematic Optimization of Long-Range Corrected
  Hybrid Density Functionals. \emph{J. Chem. Phys.} \textbf{2008}, \emph{128},
  084106\relax
\mciteBstWouldAddEndPuncttrue
\mciteSetBstMidEndSepPunct{\mcitedefaultmidpunct}
{\mcitedefaultendpunct}{\mcitedefaultseppunct}\relax
\EndOfBibitem
\bibitem[Windom and Bartlett(2023)Windom, and Bartlett]{Windom2023JCP}
Windom,~Z.~W.; Bartlett,~R.~J. On the Iterative Diagonalization of Matrices in
  Quantum Chemistry: {{Reconciling}} Preconditioner Design with
  {{Brillouin}}--{{Wigner}} Perturbation Theory. \emph{J. Chem. Phys.}
  \textbf{2023}, \emph{158}, 134107\relax
\mciteBstWouldAddEndPuncttrue
\mciteSetBstMidEndSepPunct{\mcitedefaultmidpunct}
{\mcitedefaultendpunct}{\mcitedefaultseppunct}\relax
\EndOfBibitem
\bibitem[dav()]{davidsoncode}
Minimal auxiliary basis preconditioned Davidson implementation available at
  \href{https://github.com/John-zzh/Davidson}{github.com/John-zzh/Davidson}\relax
\mciteBstWouldAddEndPuncttrue
\mciteSetBstMidEndSepPunct{\mcitedefaultmidpunct}
{\mcitedefaultendpunct}{\mcitedefaultseppunct}\relax
\EndOfBibitem
\bibitem[Zhou and Parker(2023)Zhou, and Parker]{osfdata}
Zhou,~Z.; Parker,~S.~M. Minimal auxiliary basis set approach for the electronic
  excitation spectra of organic molecules; DOI: 10.17605/OSF.IO/X5BSV. 2023;
  \url{osf.io/x5bsv}\relax
\mciteBstWouldAddEndPuncttrue
\mciteSetBstMidEndSepPunct{\mcitedefaultmidpunct}
{\mcitedefaultendpunct}{\mcitedefaultseppunct}\relax
\EndOfBibitem
\end{mcitethebibliography}

\end{document}